\documentclass[12pt]{iopart}
\usepackage{epsfig}
\begin{document}

\title{Nuclear Modification to Parton Distribution Functions 
and Parton Saturation}
\author{Zhong-Bo Kang$^1$ and Jian-Wei Qiu$^{1,2}$}
\address{$^1$Department of Physics and Astronomy, Iowa State University, 
         Ames, IA 50011, USA\\
         $^2$Physics Department, Brookhaven National Laboratory,
         Upton, NY 11973, USA}
\ead{kangzb@iastate.edu {\rm and} jwq@iastate.edu}

\begin{abstract}
We introduce a generalized definition of parton distribution 
functions (PDFs) for a more consistent all-order treatment of 
power corrections.  
We present a new set of modified DGLAP evolution equations
for nuclear PDFs, and show that the resummed 
$\alpha_s A^{1/3}/Q^2$-type of leading nuclear size enhanced 
power corrections significantly slow down the growth of 
gluon density at small-$x$.  We discuss the relation between 
the calculated power corrections and the saturation phenomena.
\end{abstract}
%===================================================================
% Introduction
%===================================================================
\section{Introduction}
RHIC has produced good evidence that a new state of hot and 
dense matter of quarks and gluons, the quark-gluon plasma (QGP), 
was formed in ultra-relativistic heavy ion collisions \cite{rhic-3yrs}. 
To extract useful information on QGP properties, we need to  
understand the dynamics of parton multiple scattering 
and to derive precise nuclear PDFs (nPDFs) to calibrate 
the production of hard probes.  
In this talk, we use lepton-hadron deep inelastic scattering (DIS) as 
an example to demonstrate a need to modify the definition of PDFs for 
all-order treatments of parton multiple scattering and 
power corrections \cite{kq-dis}.
We present a new set of parton evolution equations consistent to
QCD factorization beyond leading power and show that 
one-loop $\alpha_s A^{1/3}/Q^2$-type of power corrections
significantly slow down the growth of gluon density 
at small-$x$.  We argue that power corrections are very important 
for understanding the transition from a region where 
leading power pQCD has been very successful to 
the regime of parton saturation  
where new and novel strong interaction phenomena emerge
\cite{Mueller:1999wm}.

%===================================================================
% Factorization and parton evolution
%===================================================================
\section{QCD factorization and parton distribution functions}

\begin{figure}[ht]
\begin{minipage}[c]{1.15in}
\includegraphics[width=1.15in]{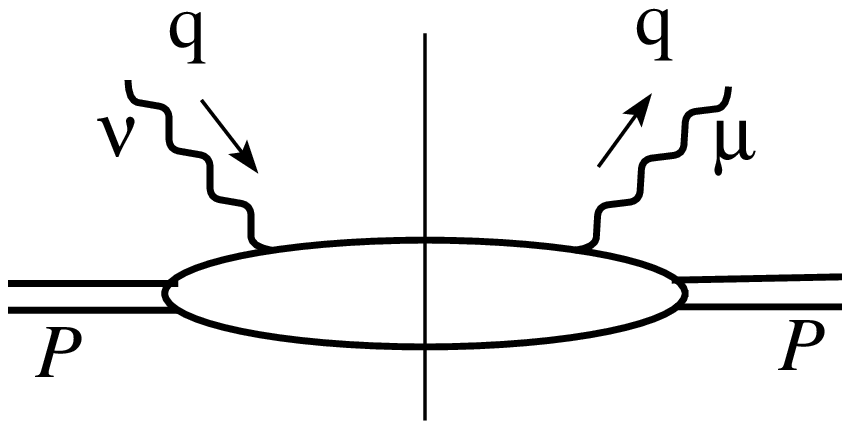}
\end{minipage}
$\approx$
\begin{minipage}[c]{0.92in}
\includegraphics[width=0.92in]{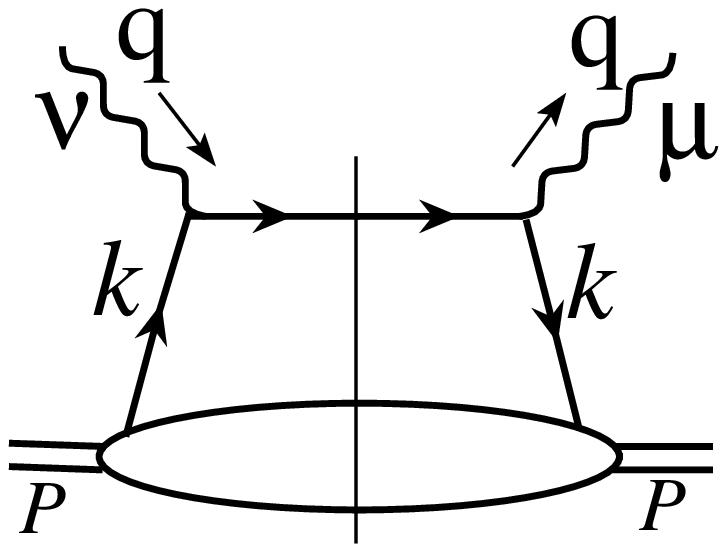}
\end{minipage}
 + 
\begin{minipage}[c]{0.92in}
\includegraphics[width=0.92in]{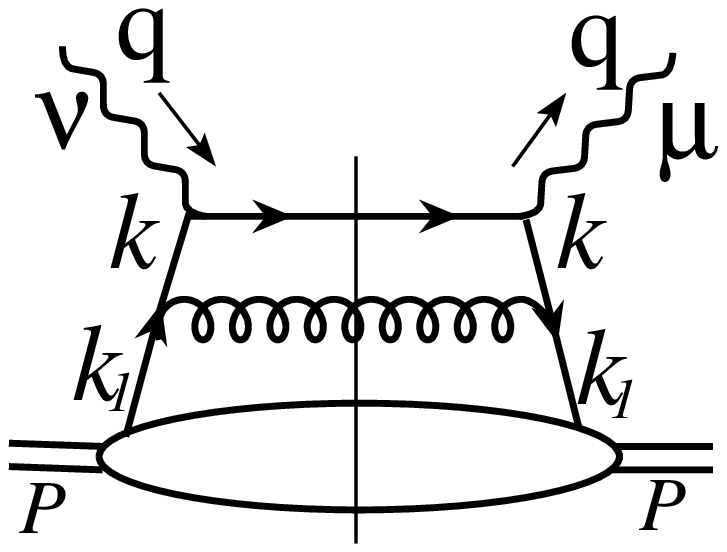}
\end{minipage}
 + 
\begin{minipage}[c]{0.92in}
\includegraphics[width=0.92in]{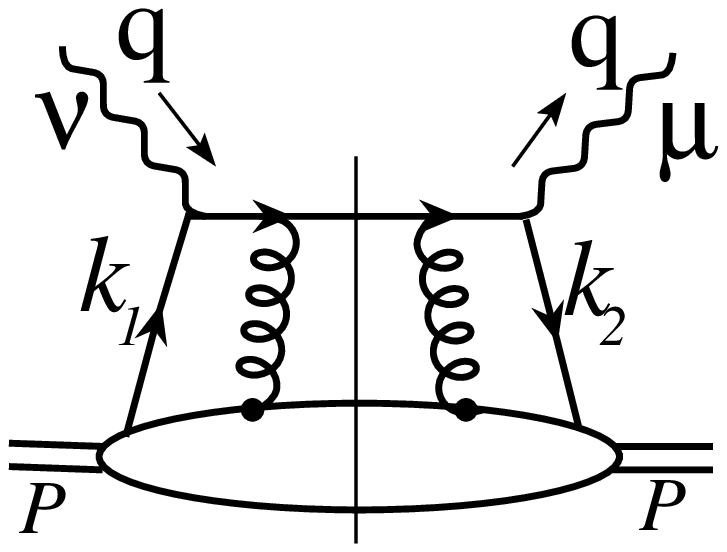}
\end{minipage}
 + 
\begin{minipage}[c]{0.92in}
\includegraphics[width=0.92in]{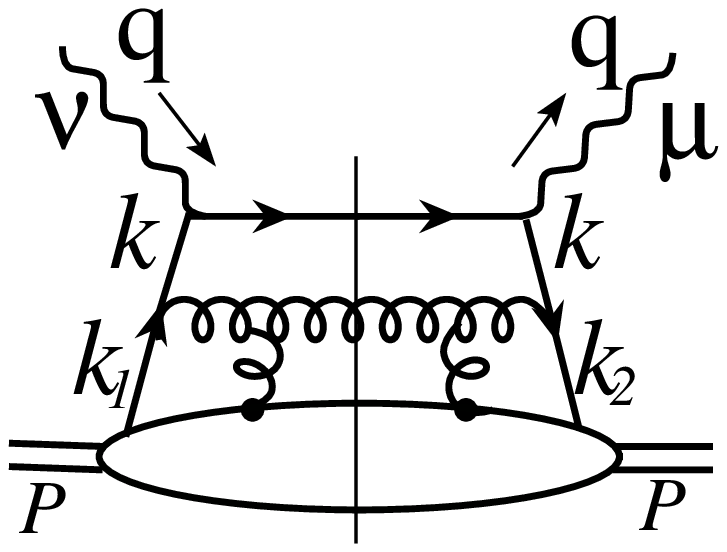}
\end{minipage}
 + $\dots$
\caption{Feynman diagram expansion of DIS hadronic tensor 
$W_{\mu\nu}$.}
\label{dis-wmn}
\end{figure}

\begin{figure}[ht]
$
\begin{minipage}[c]{0.9in}
\includegraphics[width=0.9in]{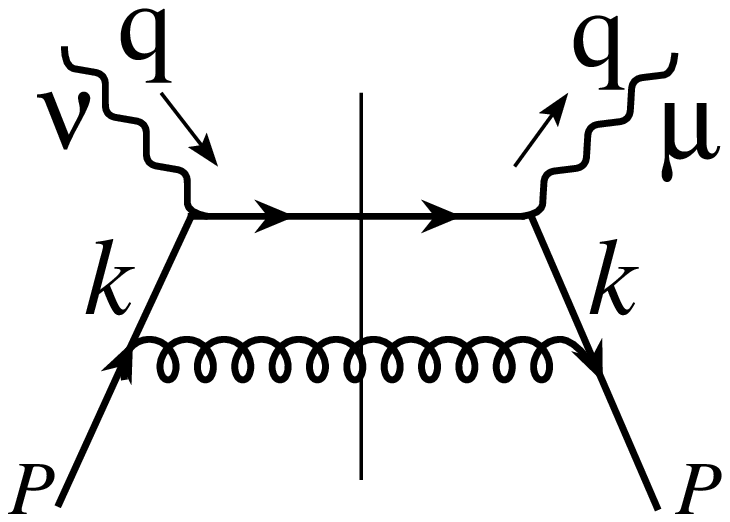}
\end{minipage}
+\dots
=H^{\mu\nu(1)}_{2(q)}\left(x_B,\frac{Q^2}{\mu^2}\right)
+ 
\begin{minipage}[c]{0.85in}
\includegraphics[width=0.85in]{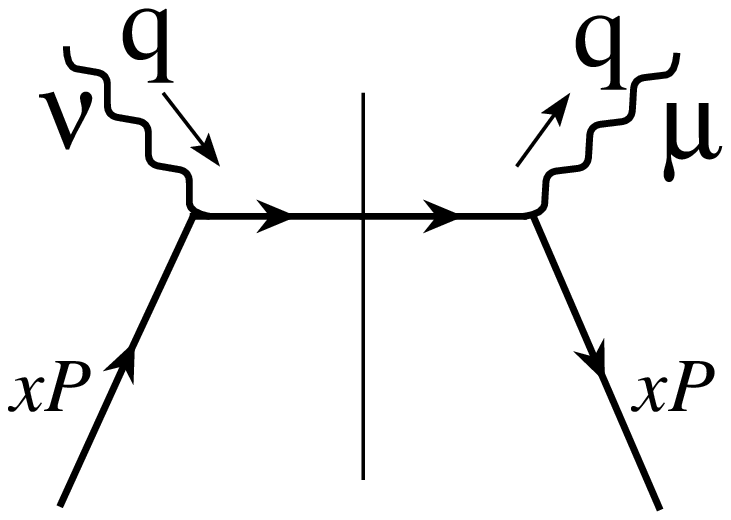}
\end{minipage}
\bigotimes
\left[
\begin{minipage}[c]{0.8in}
\includegraphics[width=0.8in]{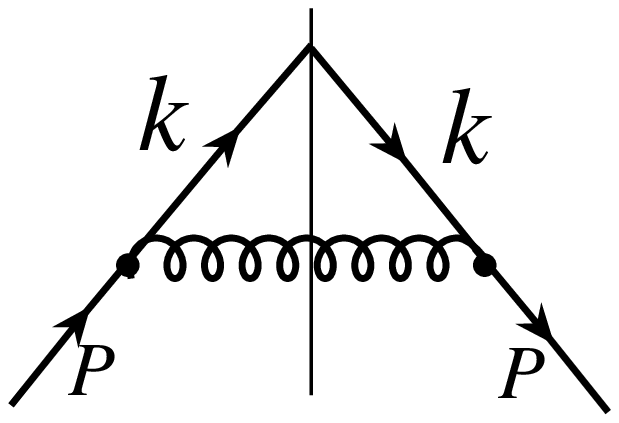}
\end{minipage}
+\dots +\mbox{UV c.t.}(\mu^2)
\right]
$
\caption{Sample calculation of leading power partonic tensor at NLO.}
\label{lt-nlo-q}
\end{figure}

Under the one-photon approximation, DIS cross section is determined by 
the hadronic tensor, $W^{\mu\nu}(P,q)$,
of hadron momentum $P$ and virtual photon momentum $q$, 
represented by the imaginary part of partonic 
scattering diagrams in Fig.~\ref{dis-wmn}.  
When $Q^2=-q^2$ is much greater than the hadron size 1/fm, 
diagrams in Fig.~\ref{dis-wmn} are dominated by the 
phase space near $k^2\sim (1/{\rm fm})^2 \ll Q^2$ 
due to the perturbative pinch singularity at $k^2\sim 0$ 
\cite{qiu-qm2002}.
By applying the operator product expansion, 
the hadronic tensor can be systematically expanded 
in a power series of $1/Q^2$ and perturbatively factorized as 
\cite{Curci:1980uw,qs-fac},
\begin{equation}
W^{\mu\nu}(x_B,Q^2) 
= \sum_{n=0}\left(\frac{1}{Q^2}\right)^n
    {\cal H}^{\mu\nu}_{2+2n}
        \left(\frac{x_B}{\{x_i\}},\frac{Q^2}{\mu^2}\right) 
    \otimes f_{2+2n}(\{x_i\},\mu^2) 
\label{wmn-fac}
\end{equation}
where $x_B=Q^2/2P\cdot q$, $\otimes$ represents the
convolution of parton momentum fractions $x_i$ with $i=1+2n$,
$f_2$'s are PDFs, $f_i$'s with $i\ge 4$ are  
twist-$i$ parton correlation functions (PCFs),
and ${\cal H}_i^{\mu\nu}$ are corresponding photon-parton scattering 
amplitudes with all perturbative collinear divergences removed.  
In Eq.~(\ref{wmn-fac}) and below, we suppress parton flavor dependence. 
The ${\cal H}_i^{\mu\nu}$ can be perturbatively computed 
order-by-order in powers of $\alpha_s$ 
by applying Eq.~(\ref{wmn-fac}) to a partonic state.
At leading power, 
tree-level quark scattering amplitude provides
parton model formalism.  
Radiative corrections at ${\cal O}(\alpha_s)$ can be derived,
for example, in Fig.~\ref{lt-nlo-q}, by 
calculating the ${\cal O}(\alpha_s)$ quark scattering amplitude 
on the left and 
${\cal O}(\alpha_s)$ quark distribution on the right.
QCD factorization assures that perturbative collinear divergence of 
the quark scattering amplitude on the left is exactly canceled by 
the same divergence of quark distribution on the right, and 
${\cal H}_{2(q)}^{\mu\nu}$ is infrared safe.  
The virtuality of the loop momentum $k$ of the scattering amplitude 
on the left is limited by the kinematics at ${\cal O}(Q^2)$. 
The $k$ in quark distribution on the right leads to a UV 
divergence that is removed by imposing a counterterm 
at a factorization scale $\mu^2\sim {\cal O}(Q^2)$. 
From Fig.~\ref{lt-nlo-q}, 
all low mass states of the quark scattering amplitude 
at $|k^2|\le \mu^2$ are canceled by that of the quark distribution
on the right, and 
${\cal H}^{\mu\nu}_{2(q)}$ 
keeps only the short-distance scattering dynamics at ${\cal O}(Q^2)$. 
The $\mu^2$ dependence 
of ${\cal H}^{\mu\nu}_{2(q)}$ is an
immediate consequence of the definition of parton-level
quark distribution in Fig.~\ref{lt-nlo-q}, and 
leads to DGLAP evolution equations of the PDFs. %\cite{DGLAP}.

%===================================================================
% Power corrections and modified parton evolution
%===================================================================
\section{Power corrections and modified parton evolution}

\begin{figure}[ht]
$
\begin{minipage}[c]{0.9in}
\includegraphics[width=0.9in]{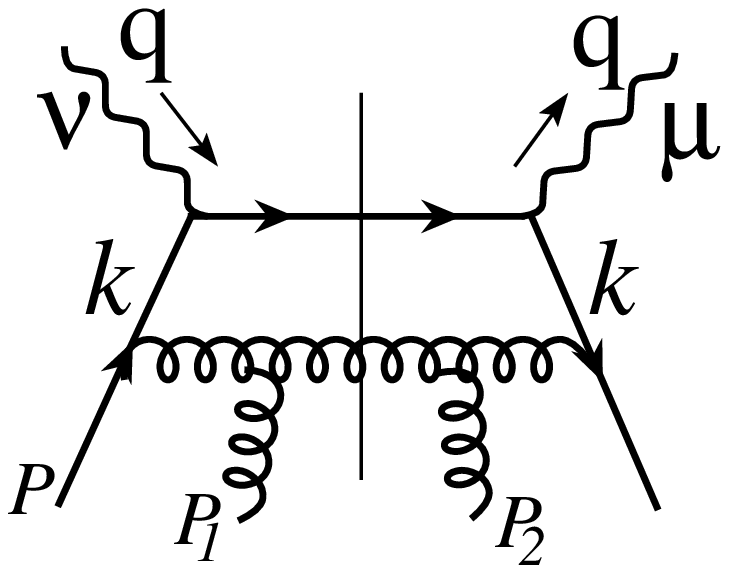}
\end{minipage}
+\dots
={\cal H}^{\mu\nu(2)}_{4(qg)}\left(x_B,\frac{Q^2}{\mu^2}\right)
+ 
\begin{minipage}[c]{0.85in}
\includegraphics[width=0.85in]{dis-wmu-LT-LO-q.ps}
\end{minipage}
\bigotimes
\left[
\begin{minipage}[c]{0.8in}
\includegraphics[width=0.8in]{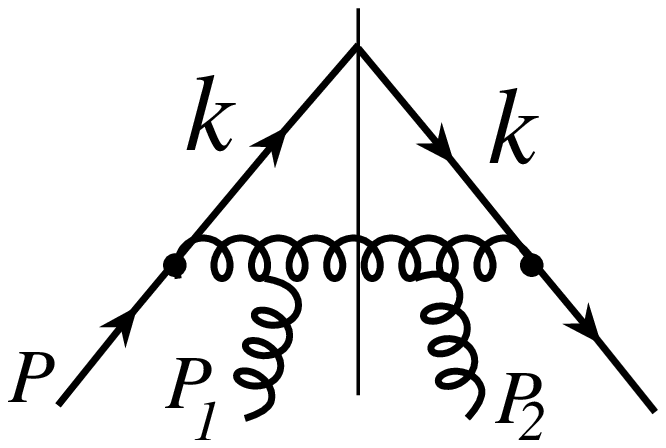}
\end{minipage}  
+\dots\
\right]
$
\caption{Sample calculation of $1/Q^2$ power corrections at NLO.}
\label{nlt-nlo-qg}
\end{figure}

To expand the reach of pQCD and explore the transition region to 
nonperturbative physics, systematic and reliable calculations of 
power corrections in Eq.~(\ref{wmn-fac}) are crucial.  
When $x_B\ll x_c=1/2mR\sim 0.1$ with hadron mass $m$ and radius $R$, 
the virtual photon in DIS could cover whole hadron
in longitudinal direction to enhance power corrections 
\cite{qiu-qm2002}. 
Power corrections from tree-level diagrams were recently computed 
and found to be important at small $x_B$ \cite{qv-prl}. 
The ${\cal O}(\alpha_s)$ correction could be computed 
in a way similar to the leading power. 
For example, in Fig.~\ref{nlt-nlo-qg}, 
${\cal H}_{4(qg)}^{\mu\nu(2)}$ is infrared safe with 
all collinear divergence canceled between the left and the right.
However, unlike the leading power case in Fig.~\ref{lt-nlo-q},
the $k$-integration of the diagram for the quark 
distribution on the right is UV finite, and consequently, 
${\cal H}_{4(qg)}^{\mu\nu(2)}$ contains the physics of 
high mass states ($|k^2|>Q^2$)   
that were not in the partonic scattering amplitude on the left.

Physically, we want {\it all} 
short-distance coefficient functions in QCD factorization formulas, 
like ${\cal H}_i^{\mu\nu}$ in Eq.~(\ref{wmn-fac}), to 
contain only dynamics of partonic scattering at ${\cal O}(Q^2)$.  
We can achieve this by defining the PDFs in Fig.~\ref{nlt-nlo-qg} 
to contain only mass states with $|k^2| < \mu^2 \sim Q^2$.  
Consequently, from this additional $\mu^2$ dependence of PDFs, 
DGLAP parton evolution equations receive power corrections 
\cite{kq-dis,mq-npb}
\begin{equation}
\mu^2\frac{\partial}{\partial\mu^2}
f_2(x, \mu^2)=\sum_{n=0}\left(\frac{1}{\mu^2}\right)^n
               {\cal P}_{2+2n}
               \left(\frac{x}{\{y_i\}}\right) 
               \otimes f_{2+2n}(\{y_i\},\mu^2) \, ,
\label{dglap}
\end{equation}
where $f_i$ with $i\ge 4$ obey similar equations 
to form a closed set of evolution equations \cite{kq-dis}.  
The ${\cal P}$'s can be perturbatively computed 
from $\mu^2$ dependence of PDFs and PCFs.

If the parton momentum $x \gg x_s$ with saturation scale 
$Q_s^2(x_s) = \mu^2$ \cite{Mueller:1999wm}, the one-loop 
contribution to quark distribution in Fig.~\ref{nlt-nlo-qg} 
is dominated by diagrams with two un-pinched poles that 
fix two of the three parton momentum integrals.
Under this leading pole approximation \cite{qs-hard}, 
only diagrams with maximum un-pinched poles contribute 
and all $y_i$ but one in Eq.~(\ref{dglap}) are fixed by these poles.
We computed all one-loop leading $\alpha_s A^{1/3}/Q^2$-type of 
power corrections to parton evolution of nPDFs \cite{kq-dis}
\begin{eqnarray}
&~&\mu^2\frac{\partial}{\partial{\mu^2}}
\left(\begin{array}{c} q(x,\mu^2) \\ g(x,\mu^2) \end{array} \right)
=\frac{\alpha_s(\mu^2)}{2\pi}\int_x^1\frac{dy}{y}
\nonumber\\
&~&\times\left(\begin{array}{cc} P_{qq}(\frac{x}{y}){\hat T}_{\Delta'} 
& P_{qg}(\frac{x}{y}){\hat T}_{\Delta} \\
\sum_{i=1}^{2n_f}P_{gq}(\frac{x}{y}){\hat T}_{\Delta} 
& P_{gg}(\frac{x}{y}){\hat T}_{\Delta'}
-\frac{n_f}{3}\delta(1-\frac{x}{y}){\hat T}_{\Delta}
\end{array}\right)
\left(\begin{array}{c} q(y,\mu^2) \\ g(y,\mu^2) \end{array} \right)
\label{mdglap}
\end{eqnarray}
where ${\hat T}_{\Delta}$ is a translation operator that shifts
momentum fraction $y\to y(1+\Delta)$ with
$\Delta=\frac{C_F}{C_A}\Delta'=\xi^2(A^{1/3}-1)/Q^2$ and $\xi^2$
defined in \cite{qv-prl}, and 
$P_{qq},P_{qg},P_{gq}$ are normal leading order 
splitting functions, while $P_{gg}$ is the normal $g\to g$
splitting function with the term proportional to $n_f$ excluded. 
\begin{figure}[h!]
\begin{center}
\psfig{file=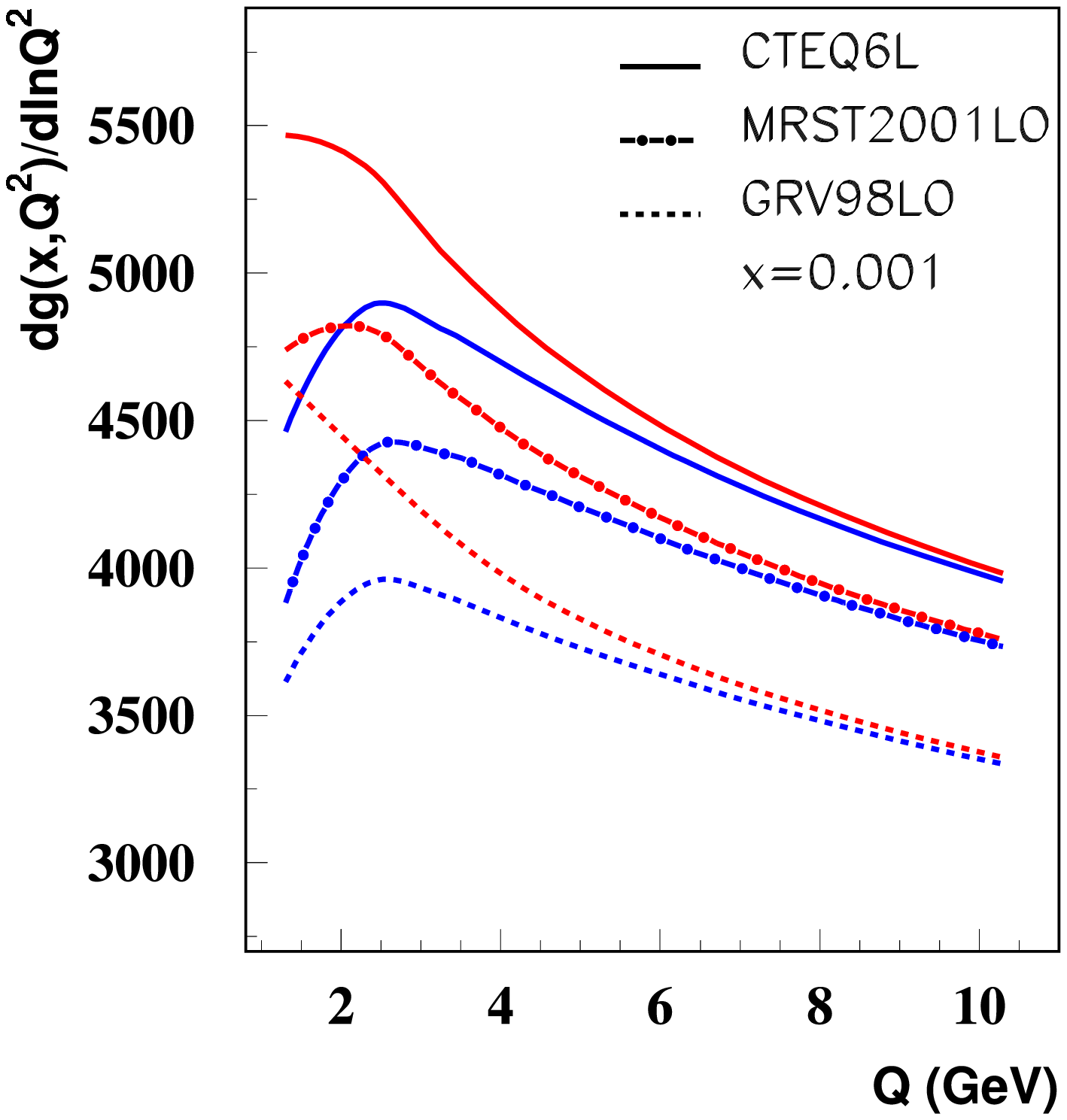,height=2.0in,width=2.2in}
\hskip 0.2in
\psfig{file=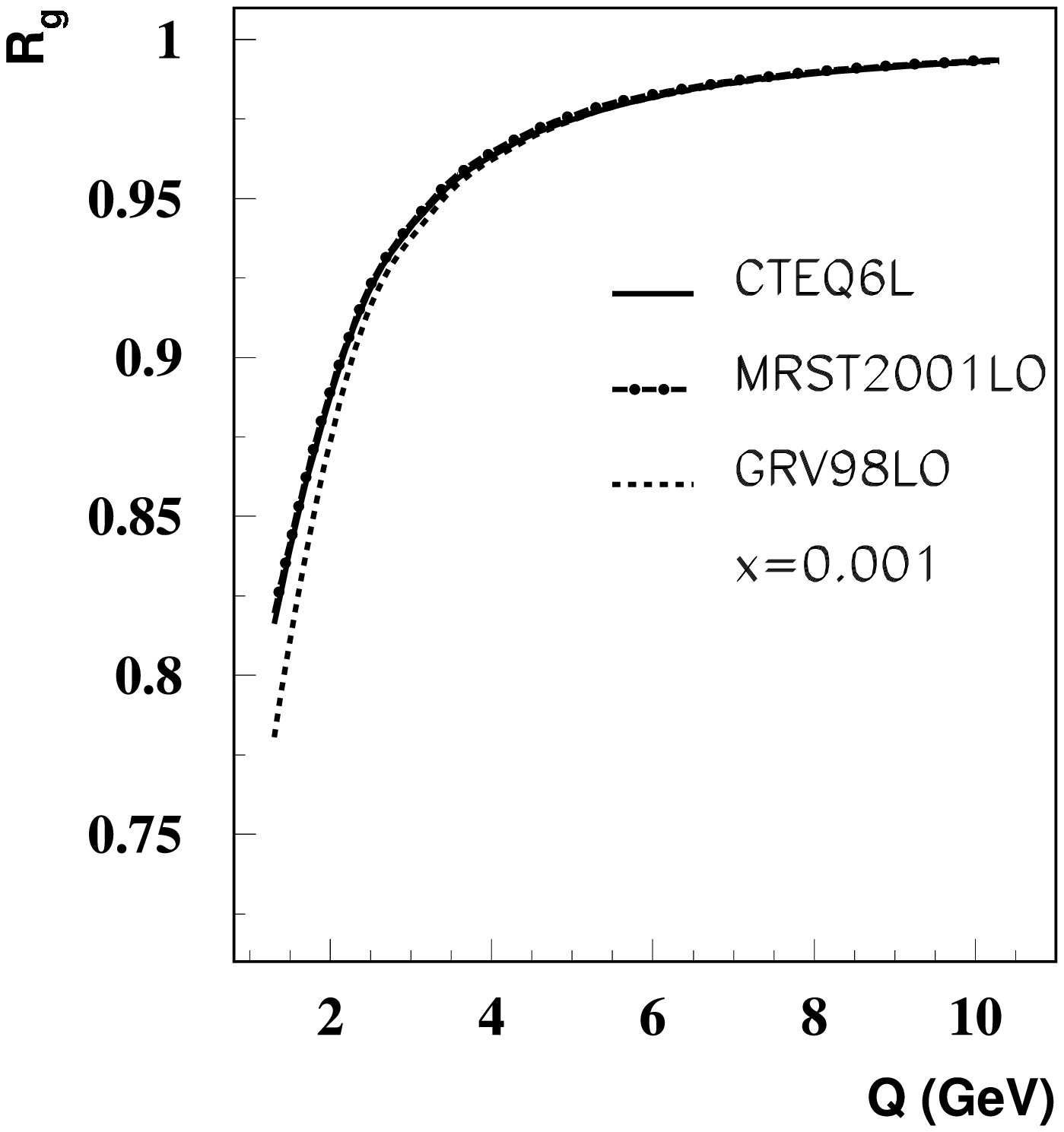,height=2.0in,width=2.2in}

\vskip -0.2in
\hskip 0.5in (a)\hskip 2.2in (b)

\vskip -0.2in
\end{center}
\caption{Modification of gluon evolution for nucleus of $A=197$}
\label{slope}
\end{figure}
In Fig.~\ref{slope}(a) and (b), we show gluon evolution without (red) 
and with (blue) power corrections, and the ratio of these two 
different evolutions, respectively.  
Using three sets of available PDFs, we found 
that the resummed power corrections can give as large as 
20\% reduction in {\it slope} of gluon evolution and can significantly 
slow down the growth of gluon density at small-$x$ and low $Q^2$. 

%===================================================================
% summary and outlook
%===================================================================
\section{Summary and outlook}

We modified the definition of PDFs so that all 
short-distance coefficient functions of power corrections 
in QCD factorization formulas contain only dynamics of 
partonic scattering at ${\cal O}(Q^2)$.  Consequently, 
we obtained a new set of parton evolution equations including
calculable power corrections in Eq.~(\ref{dglap}).
Under the leading pole approximation, we computed all one-loop 
contributions to the evolution equations of nPDFs and found that   
leading $\alpha_s A^{1/3}/Q^2$-type power corrections 
significantly slow down the growth of gluon density 
at small-$x$ and drive the gluon density to saturation \cite{kq-dis}.

Our factorization treatment of power corrections
can be applied to any factorizable observables. 
Power corrections closely connect to coherent 
multi-parton dynamics and are important for  
studying the transition from a 
dilute to a saturated partonic system.
When partons are close to be saturated, or
all $x_i$ are close to the $x_s$, the leading pole approximation 
is no longer reliable.  Many more diagrams will contribute. 
We need different approaches for studying power corrections 
and parton correlations \cite{Mueller:1999wm,mq-npb}.

%===================================================================
We acknowledge support from the U.S. Department of Energy 
Grant No. DE-FG02-87ER40371 and Contract No. DE-AC02-98CH10886.

%===================================================================
% references
%===================================================================

%===================================================================
\end{document}